\documentclass[aps,twocolumn,preprintnumbers,superscriptaddress,floatfix,prl,10pt]{revtex4-2}
\usepackage{bbm}
\usepackage{bm}
\usepackage{mathrsfs }
\usepackage{amsmath}
\usepackage{amssymb}
\usepackage{empheq}
\usepackage{graphicx}
\usepackage{mathrsfs}
\usepackage{amsfonts}
\usepackage{amsthm}
\usepackage{color}
\usepackage{multirow}
\usepackage{bigints}
\usepackage{txfonts}
\usepackage{float}
\usepackage{hyperref}
\hypersetup{
     unicode=false,              % non-Latin characters in Acrobat?s bookmarks
     pdftoolbar=true,           % show Acrobat?s toolbar?
     pdfmenubar=true,         % show Acrobat?s menu?
     pdffitwindow=false,      % window fit to page when opened
     pdfstartview={FitH},    % fits the width of the page to the window
     pdftitle={My title},       % title
     pdfauthor={Author},     % author
     pdfsubject={Subject},   % subject of the document
     pdfcreator={Creator},   % creator of the document
     pdfproducer={Producer}, % producer of the document
     pdfkeywords={keyword1} {key2} {key3}, % list of keywords
     pdfnewwindow=true, % links in new PDF window
     colorlinks=false,       % false: boxed links; true: colored links
     linkcolor=red,           % color of internal links (change box color with linkbordercolor)
     citecolor=green,        % color of links to bibliography
     filecolor=magenta,    % color of file links
     urlcolor=cyan           % color of external links
}

\makeatletter \tolerance = 10000 \tolerance = 10000

\makeatother
\begin{document}

\title{Unified one-parameter scaling function for Anderson localization transitions in  non-reciprocal non-Hermitian systems}

\author{C. Wang}
\email[Corresponding author: ]{physcwang@tju.edu.cn}
\affiliation{Center for Joint Quantum Studies and Department of Physics, School of Science, Tianjin University, Tianjin 300350, China}
\author{Wenxue He}
\affiliation{Center for Joint Quantum Studies and Department of Physics, School of Science, Tianjin University, Tianjin 300350, China}
\affiliation{Tianjin Key Laboratory of Low Dimensional Materials Physics and Preparing Technology, School of Science, Tianjin University, Tianjin 300072}
\author{X. R. Wang}
\affiliation{Physics Department, The Hong Kong University of Science 
and Technology (HKUST), Clear Water Bay, Kowloon, Hong Kong}
\affiliation{HKUST Shenzhen Research Institute, Shenzhen 518057, China}
\author{Hechen Ren}
\email[Corresponding author: ]{ren@tju.edu.cn}
\affiliation{Center for Joint Quantum Studies and Department of Physics, School of Science, Tianjin University, Tianjin 300350, China}
\affiliation{Tianjin Key Laboratory of Low Dimensional Materials Physics and Preparing Technology, School of Science, Tianjin University, Tianjin 300072}
\affiliation{Joint School of National University of Singapore and Tianjin University, International Campus of Tianjin University, Binhai New City, Fuzhou 350207, China}

\date{\today}

\begin{abstract}
%By using dimensionless conductances as scaling variables, the conventional one-parameter scaling theory of localization predicts that all states in disordered one-dimensional Hermitian systems are localized. However, this paradigm
By using dimensionless conductances as scaling variables, the conventional one-parameter scaling theory of localization fails for non-reciprocal non-Hermitian systems such as the Hanato-Nelson model. Here, we propose a one-parameter scaling function using the participation ratio as the scaling variable. Employing a highly accurate numerical procedure based on exact diagonalization, we demonstrate that this one-parameter scaling function can describe Anderson localization transitions of non-reciprocal non-Hermitian systems in one and two dimensions of symmetry classes AI and A. The critical exponents of correlation lengths depend on symmetries and dimensionality only, a typical feature of universality. Moreover, we derive a complex-gap equation based on the self-consistent Born approximation that can determine the disorder at which the point gap closes. The obtained disorders match perfectly the critical disorders of Anderson localization transitions from the one-parameter scaling function. Finally, we show that the one-parameter scaling function is also valid for Anderson localization transitions in reciprocal non-Hermitian systems such as two-dimensional class AII$^\dagger$ and can, thus, serve as a unified scaling function for disordered non-Hermitian systems.
\end{abstract}

\maketitle

\emph{Introduction.~}The orthodox one-parameter scaling theory (1PST) of localization has been instrumental in understanding Anderson localization transitions (ALTs) in disordered Hermitian systems~\cite{jtEdwards_jpc_1972,dcLicciardello_jpc_1975,eAbrahams_prl_1979,paLee_rmp_1985,bKramer_rpp_1993,bHuckestein_rmp_1995,fEvers_rmp_2008}. This theory assumes that the scaling function $\beta(g)$, describing how the dimensionless conductance $g$ changes with system size, is a univariate function of $g$ and predicts that symmetries and dimensionality of disordered systems determine the universality class of localization. According to the 1PST, ALTs can (cannot) happen in three-dimensional (one-dimensional, 1D) disordered Hermitian systems~\cite{eAbrahams_prl_1979}, while, in two dimensions (2D),  the occurrence of  ALTs depends on the combination of time-reversal and spin-rotational symmetries~\cite{snEvangelou_prl_1995,rMerkt_prb_1998,yasada_prl_2002,pMarkos_jpa_2006,cWang_prl_2015,ySu_scirep_2016,ySu_scirep_2016,gOrso_prl_2017,cWang_prb_2017,wChen_prb_2019}.
\par

\begin{figure}[htbp]
\centering
\includegraphics[width=0.48\textwidth]{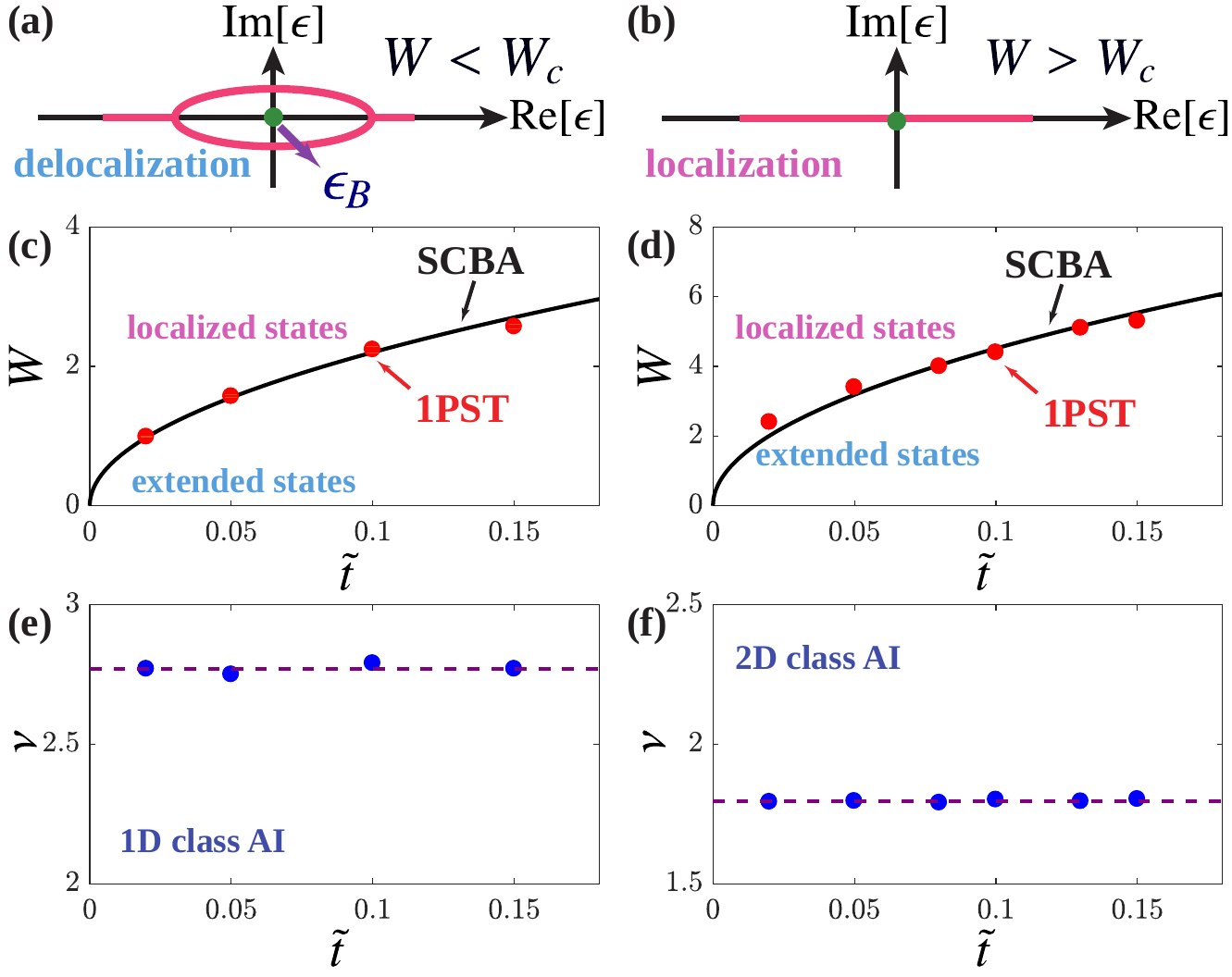}
\caption{(a) Schematic plot of the energy spectrum of a 1D non-reciprocal system with point gap $\epsilon_{\text{B}}=0$ and $W<W_c$ where the $\text{Re}[\epsilon]=0$ states are extended states. (b) Energy spectrum of the same system for $W>W_c$ where the $\text{Re}[\epsilon]=0$ state is localized and the point-gap closes. (c) Phase diagram on the $W-\tilde{t}$ plane of $H^{d=1}_{\text{AI}}$. The phase boundary (black line) is determined by the SCBA, and the red circles are obtained by using the one-parameter scaling function Eq.~\eqref{eq_2}. (d) Same as (c) but for $H^{d=2}_{\text{AI}}$. (e) Critical exponents $\nu$ as a function $\tilde{t}$ for $H^{d=1}_{\text{AI}}$. Dashed line locates the average $\nu=2.77$. (f) Same as (e) but for  $H^{d=2}_{\text{AI}}$. Dashed line marks the average critical exponent $\nu=1.79$.}
\label{fig1}
\end{figure}

Recently, non-Hermitian systems have attracted great interest because of their exotic properties, such as the generalized bulk-boundary correspondence~\cite{teLee_prl_2016,hShen_prl_2018,fkKunst_prl_2018,sYao_prl_2018,tLiu_prl_2019}, parity-time symmetry~\cite{cmBender_prl_1998,ceRuter_natphys_2010,bPeng_natphys_2014,lFeng_science_2014,vvKonotop_rmp_2016,lXiao_natphys_2017,sWeimann_natmater_2017,hYang_prl_2018}, and exceptional points~\cite{hXu_nature_2016,bPeng_PNAS_2016,yXu_prl_2017,zXiao_prl_2019,skOzdemir_natmat_2019,maMiri_science_2019,xxZhang_prl_2020,kBai_prl_2023}. As most physical systems realizing such models possess disorders, ALTs in non-Hermitian systems have emerged as a research focus among both theoretical and experimental communities~\cite{nHatano_prl_1996,rHamazaki_prl_2019,yHuang_prb_2020,cWang_prb_2020,ljZhai_prb_2020,xLuo_prl_2021,kKawabata_prl_2021,xLuo_prb_2021,xLuo_prresearch_2022,kSuthar_prb_2022,cWang_prb_2022,cWang_prb_2023,cWang_prb_2024}. Early studies in this area have presented surprising findings suggesting uncommon localization phenomena. For example, while orthodox 1PST says no extended states in 1D~\cite{eAbrahams_prl_1979}, ALTs can appear in 1D disordered non-Hermitian systems with asymmetric hooping as first predicted by Hatano and Nelson~\cite{nHatano_prl_1996}. Such non-reciprocal systems display a point gap with respect to a specific point $\epsilon_{\text{B}}$ on the complex energy plane and become an Anderson insulator at strong disorders when the point gap closes~\cite{zGong_prx_2018,kKawabata_prx_2019}; see Figs.~\ref{fig1}(a) and (b). 
\par

Some non-Hermitian systems are reciprocal, where only gain or loss are involved and 1PST is still valid~\cite{xLuo_prb_2021}. Conversely, to describe the ALTs in 1D disordered non-reciprocal non-Hermitian systems like Hatano-Nelson model~\cite{nHatano_prl_1996}, a two-parameter scaling theory is suggested~\cite{kKawabata_prl_2021}, where two conductances $g_{\text{L}}$ and $g_{\text{R}}$ in opposite transport directions appear in the scaling function. Therefore, a unified scaling theory for ALTs in non-Hermitian systems is lacking, especially for higher dimensions, and whether a one-parameter scaling function can be applied to ALTs of non-reciprocal non-Hermitian systems remains an open question.
\par

Here, we develop a novel 1PST for ALTs of non-reciprocal non-Hermitian systems, which can also be applied to reciprocal non-Hermitian systems as substantiated here and in our early works~\cite{cWang_prb_2020,cWang_prb_2022,cWang_prb_2023}. Rather than dimensionless conductances, we use the participation ratio of periodic systems as the scaling variable. Through the scaling function, one can determine the critical disorders $W_c$ and the critical exponents $\nu$ of correlation lengths; see two representative examples shown in Figs.~\ref{fig1}(c-f) for 1D and 2D class AI according to the 38-fold classification~\cite{kKawabata_prx_2019}. We further derive a complex-gap equation based on the self-consistent Born approximation (SCBA) that can predict the critical disorder $W_c$ at which the point gap closes~\cite{pSheng_book_2006}. The obtained $W_c$ matches that from our one-parameter scaling function. Finally, our scaling theory predicts additional universality classes in 1D and 2D non-Hermitian systems with different symmetries, either reciprocal or non-reciprocal; see Table~\ref{table_1}.  
\par 

\begin{table}
\caption{Critical exponents obtained by the scaling function Eq.~\eqref{eq_2} of ALTs in disordered non-Hermitian systems with (without) reciprocity for different symmetries and dimensionality.}
\begin{ruledtabular}
\begin{tabular}{ccccc}
symmetry class & dimension & reciprocity & $\nu$ & $Q$ \\
\hline
AI   & 1 & no  & $2.77\pm 0.06$  & 0.2   \\
AI   & 2 & no  & $1.88\pm 0.03$  & 0.09 \\
A     & 1 & no  & $2.2\pm 0.3$      & 0.07 \\
A     & 2 & no  & $2.06\pm 0.03$  & 0.1    \\
AII$^\dagger$  & 2 & yes & $1.39\pm 0.05$  & 0.09 \\
AIII & 2 & yes & $2.0\pm 0.1$~\cite{cWang_prb_2023}  & 0.1 \\
DIII+$\mathcal{S}_{-+}$ & 2 & yes & $2.0\pm 0.1$~\cite{cWang_prb_2023} & 0.08 \\
\end{tabular}
\end{ruledtabular}\label{table_1}
\end{table}

\emph{1D Class AI.}~To introduce the one-parameter scaling function, we first focus on the generalized Hanato-Nelson model of volume $V=L^d$ and lattice constant $a=1$~\cite{nHatano_prl_1996}. The Hamiltonian of the system is
\begin{equation}
\begin{gathered}
H^{d}_{\text{AI}}=\sum_{\bm{i}}c^\dagger_{\bm{i}}\epsilon_{\bm{i}}c_{\bm{i}}+\sum_{\langle \bm{ij}\rangle}c^\dagger_{\bm{i}}(t+\tilde{t})c_{\bm{j}}+c^\dagger_{\bm{j}}(t-\tilde{t})c_{\bm{i}}
\end{gathered}\label{eq_1}
\end{equation}
with $c_{\bm{i}}$ and $c^\dagger_{\bm{i}}$ being the single-particle annihilation and creation operators on a $d$-dimensional lattice site $\bm{i}$. $\epsilon_{\bm{i}}$ distributes uniformly in a range of $[-W/2,W/2]$ such that $W$ measures the degree of randomness. $t$ and $\tilde{t}$ are real positive numbers, and $\langle \bm{ij}\rangle$ represents the nearest-neighbor sites.  Hereafter, $t=1$ is chosen as the energy unit. Model~\eqref{eq_1} preserves time-reversal symmetry and belongs to class AI~\cite{kKawabata_prx_2019}. 
\par 

Let us first consider 1D cases. In the absence of disorder, Eq.~\eqref{eq_1} can be diagonalized in the momentum space as $H^{d=1}_{\text{AI}}=\sum_k c^\dagger_k h^{d=1}_{\text{AI}}(k) c_k$ with $h^{d=1}_{\text{AI}}(k)=2t\cos [k]+i2\tilde{t} \sin [k]$. The $\bm{k}\cdot \bm{p}$ continuous Hamiltonian near $k=\pi/2$ reads $h^{d=1}_{\text{AI}}(p)=-2tp+i\tilde{t}(2-p^2)+o(p^3)$. The non-Hermitian terms $i2\tilde{t}$ acting on the eigenstates of the Hermitian part adds a pre-factor of $e^{2\tilde{t} x}$ to the Bloch wavefunctions. Thus, for finite-size systems with open boundary conditions, the non-Hermitian potential causes the Bloch states with well-defined momenta to exponentially localize at sample boundaries~\cite{cWang_prb_2022_2,cWang_prb_2022_3}. This notable phenomenology is referred to as the non-Hermitian skin effect~\cite{sYao_prl_2018}.
\par

These skin modes are destructed by strong disorders,  at which all states localizes in the bulk rather than on the boundary. To visualize the transitions from skin modes to conventional localized modes, we compute the following quantity $\Lambda_{\epsilon}(L,W)=\left(\sum^{L}_{i=1}|(\psi^{\text{L}}_{\epsilon}(i))^\ast \psi^{\text{R}}_{\epsilon}(i)| i \right)^{-1}$ with $\psi^{\text{R/L}}_{\epsilon}(i)$ being the normalized wavefunction amplitude on site $i$ of the right/left eigenstate with energy $\epsilon$ and satisfying the biorthogonal relation~\cite{bi_orthogonal}. We set $\tilde{t}>0$ such that the skin mode is localized at $i=1$. 
% Effectively, $\Lambda^{-1}_{\epsilon}$ is the distance between the wavefunction tail of state of energy $\epsilon$ and the end of $i=1$. 
If the mode is a skin mode, its wavefunction should be highly localized at $i=1$ such that $\Lambda_{\epsilon}$ is large and size-independent. Conversely, we expect $\Lambda_{\epsilon}$ is small and decrease with size $L$ for localized states that are not skin modes.
\par

\begin{figure}[htbp]
\centering
\includegraphics[width=0.48\textwidth]{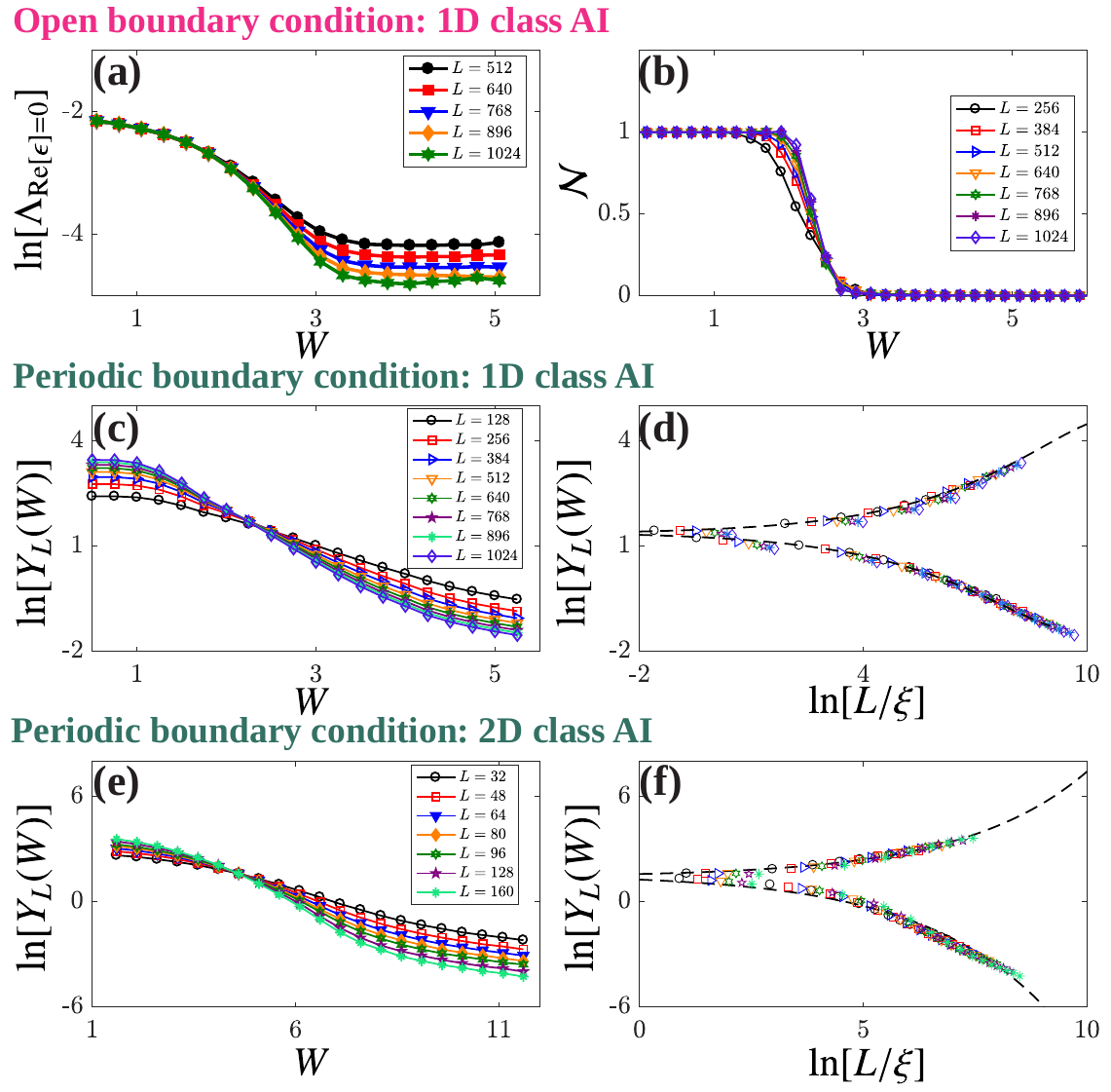}
\caption{\textbf{1D Class AI}:~(a) $\ln [\Lambda_{\text{Re}[\epsilon]=0}(L,W)]$ as a function of $W$ for $\tilde{t}=0.1$ and various $L$. (b) Winding number $\mathcal{N}$ as a function of $W$ for $\tilde{t}=0.1$ and $L=256,384,\cdots,1024$.  (c) $\ln [Y_L(W)]$ as a function $W$ for various $L=128,256,\cdots,1024$. (d) Scaling function $\ln[Y_L(W)]=\ln [f(\ln [L/\xi])]$ with $\xi\propto |W-W_c|^{-\nu}$ and $\nu=2.77\pm 0.06$. \textbf{2D Class AI}:~(e) $\ln [Y_L(W)]$ for $\text{Re}[\epsilon]=0$, $\tilde{t}=0.1$, and $L=32,48,\cdots,160$. (f) Scaling function $\ln[Y_L]=\ln [f(x=\ln[L/\xi])]$. Scaling analysis based on Eq.~\eqref{eq_2} gives $W_c=4.5\pm 0.2$ and $\nu=1.88\pm 0.03$.}
\label{fig2}
\end{figure}

Figure~\ref{fig2}(a) shows $\ln[\Lambda_{\text{Re}[\epsilon]=0}]$ as a function of $W$ for $\tilde{t}=0.1$ and varying $L$. The size-independent $\ln[\Lambda_{\text{Re}[\epsilon]=0}]$ indicates that skin modes appear at small disorders. There exists a critical disorder $W_c\simeq 2.4$ above which $\ln[\Lambda_{\text{Re}[\epsilon]=0}]$ decreases with $L$, indicating that the states are conventional localized states. 
\par

Besides, the skin modes can be characterized by a real-space quantized winding number $\mathcal{N}$ defined as $\mathcal{N}=\textit{Tr}\left[ \mathbb{Q}^\dagger [\mathbb{Q},\mathbb{X}] \right]$ with the unitary matrix $\mathbb{Q}$ obtained from the polar decomposition $(H^{d=1}_{\text{AI}}-\epsilon_{\text{B}})=\mathbb{QP}$ and $\textit{Tr}$ being the trace per unit volume~\cite{lJin_prb_2019,fSong_prl_2019,nOkuma_prl_2020,kZhang_prl_2020,jClaes_prb_2021}. Here, $\mathbb{P}$ is a positive semi-definite Hermitian matrix, and $\mathbb{X}$ is the coordinate operator; see Supplementary~\citep{supp}. The winding number as a function of $W$ is shown in Fig.~\ref{fig2}(b). We find that $\mathcal{N}$ loses it quantization (to 1) at $W_c=2.4$, consistent with the data in Fig.~\ref{fig2}(a).
\par

\emph{One-parameter scaling function.}~As we have established a transition from skin modes to conventional localized modes under open boundary conditions, we expect an ALT also happens at $W_c$ if periodic boundary condition is applied to avoid non-Hermitian skin modes~\cite{nHatano_prl_1996}. Since the conventional 1PST is invalid in non-reciprocal systems~\cite{kKawabata_prl_2021}, we propose a different one-parameter scaling function by choosing the participation ratio as the scaling variable. 
\par

The participation ratio is defined as $p_2(L,W) =\left( \sum^{L^d}_{\bm{i}=1} |(\psi^{\text{L}}_{\epsilon}(\bm{i}))^\ast \psi^{\text{R}}_{\epsilon}(\bm{i})|^2 \right)^{-1}$, and the central result of the present work is the following one-parameter scaling hypothesis 
\begin{equation}
\begin{gathered}
p_2(L,W)=L^D f(L/\xi)+\phi L^{-y} \tilde{f}(L/\xi)
\end{gathered}\label{eq_2}
\end{equation}
for a non-Hermitian system of volume $L^d$. Here, $D$ is the fractal dimension of wavefunction at critical point~\cite{jhPixley_prl_2015,cWang_prb_2019}. $\xi$ is the correlation length and diverges as $\xi\propto |W-W_c|^{-\nu}$ with $\nu$ being the critical exponent~\cite{cWang_prl_2015,cWang_prb_2022,cWang_prb_2023}. $f(x)$ and $\tilde{f}(x)$ are the scaling functions for the relevant and irrelevant scaling variables, respectively. $\phi$ is a constant, and $y>0$ is the exponent for irrelevant scaling variable. For large enough systems, the irrelevant scaling variable is insignificant such that $p_2(L,W)\simeq L^D f(L/\xi)$~\cite{xrWang_pra_1989}. 
\par

Once $p_2$ are obtained numerically through the exact diagonalizations~\cite{numerical,kwant,scipy}, we perform a $\chi^2$-fit of them to Eq.~\eqref{eq_2}. We calculate the goodness-of-fit $Q$ to evaluate the fit~\cite{whPress_book_1996}. In what follows, $Q>0.05$ for all fits, satisfying the minimum acceptable criterion $Q\geq 10^{-3}$; see Table~\ref{table_1}. We can thus use this one-parameter scaling function to identify ALTs as follows: (i) For extended (localized) states, the quantity $Y_L(W)=p_2 L^{-D}-\phi L^{-y-D} \tilde{f}(L/\xi)$ increases (decreases) with $L$; (ii) At critical disorder, $Y_L(W_c)$ is size-independent. 
\par

One representative example is shown in Fig.~\ref{fig2}(c), which displays $Y_L(W)$ for $\text{Re}[\epsilon]=0$ and $\tilde{t}=0.1$ (same as Figs.~\ref{fig2}(a) and \ref{fig2}(b)). Curves of different $L$ (range $L=128$ from to 1024) cross at $W_c=2.37\pm 0.05$, and those of $W<W_c$ ($W>W_c$) increases (decreases) of $L$. We also obtain a smooth scaling function $f(x)$ from Eq.~\eqref{eq_2}; see Fig.~\ref{fig2}(d). These features suggest an ALT at $W_c$. Furthermore, the established critical disorder is the same as that of the skin-localization transition in the open system as shown in Figs.~\ref{fig2}(a) and \ref{fig2}(c), which is consistent with previous results~\cite{zGong_prx_2018,kKawabata_prx_2019} and further validates the scaling function Eq.~\eqref{eq_2}. 
\par

Universality is a fundamental concept for ALTs, which claims the critical exponents like $\nu$ and $D$ depend on symmetries and dimensionality and should be independent on other model parameters~\cite{fEvers_rmp_2008}. To test whether the scaling function~\eqref{eq_2} respects universality, we calculate $\nu$ and $D$ as a function of $\tilde{t}$ (the degree of non-Hermiticity), see Fig.~\ref{fig1}(c) and Fig.~S6 in Supplementary~\cite{supp}. Our numerical results suggest that the two critical exponents are independent of $\tilde{t}$, a typical feature of universality. 
\par

\emph{Complex-gap equation.}~To further support our one-parameter scaling function, we calculate the critical disorder $W_c$ through the SCBA~\cite{pSheng_book_2006} and compare with those obtained through the scaling function. 
\par

We begin from the retarded/advanced Green's functions for a non-Hermitian Hamiltonian $H$ in the continuous limit: $G^{\text{R}}(\bm{r}',\bm{r},t)=-i\theta(t)\langle \bm{r}'|e^{-iHt}|\bm{r}\rangle$ and $G^{\text{A}}(\bm{r}',\bm{r},t)=i\theta(-t)\langle \bm{r}'|e^{-iHt}|\bm{r}\rangle$ with $\theta(t)$ being the Heaviside step function~\cite{eAkkermans_book_2007}. By performing a Laplace transform, we obtain the retarded/advanced Green's operators: $G^{\text{R,A}}(\epsilon)=1/(\epsilon-H\pm i0^+)$~\cite{supp}. 
\par

With the help of self-energy $\Sigma$, we can directly calculate the disorder-average retarded/advanced Green's operators: $G^{\text{R,A}}(\epsilon)=1/(\epsilon-H_0-\Sigma\pm i0^+)$ with $H_0$ being the Hamiltonian of the clean system, i.e., the effective Hamiltonian of disordered systems can be treated as $H=H_0+\Sigma$. As we focus on periodic systems and $\text{Re}[\epsilon]=0$, it is more convenient to use the low-energy effective Bloch Hamiltonian mentioned before, i.e., $h^{d=1}_{\text{AI}}(p)=-2tp+i\tilde{t}(2-p^2)$. Here, we use the SCBA, which successfully explains the appearance of topological Anderson insulators~\cite{jLi_prl_2009,cwGroth_prl_2009}, to calculate the self-energy $\Sigma$. By writing $\Sigma=\alpha+i\beta$ and $\epsilon=E+i\gamma$, we obtain~\cite{supp}
\begin{equation}
\begin{gathered}
\alpha=\dfrac{W^2}{24\pi}\int \mathrm{d}p \dfrac{E+2tp-\alpha}{(E+2tp-\alpha)^2+(\gamma+\tilde{t}p^2-2\tilde{t}-\beta+0^+ )^2}
\end{gathered}\label{eq_3}
\end{equation}
and
\begin{equation}
\begin{gathered}
\beta=-\dfrac{W^2}{24\pi}\int \mathrm{d}p \dfrac{\gamma+\tilde{t}p^2-2\tilde{t}-\beta+0^+}{(E+2tp-\alpha)^2+(\gamma+\tilde{t}p^2-2\tilde{t}-\beta+0^+ )^2}.
\end{gathered}\label{eq_4}
\end{equation}
For $\text{Re}[\epsilon]=E=0$, one can find that $\alpha=0$ is always the solution of Eq.~\eqref{eq_3}. Therefore, the energy gap of $\text{Re}[\epsilon]=0$ is renormalized by disorders as~\cite{supp} 
\begin{equation}
\begin{gathered}
\Delta \epsilon=|4\tilde{t}+2\beta|,
\end{gathered}\label{eq_5}
\end{equation}
where $\Delta\epsilon=4\tilde{t}$ at $W=0$ and decreases with increasing of $W$. 
\par

The complex-gap equation~\eqref{eq_5} can determine the critical disorder in Fig.~\ref{fig2}(c). Note that the mobility edge is the circle enclosing the origin of the complex energy plane shown in Fig.~\ref{fig1}(a) and shrinks to the origin with increasing $W$~\cite{teLee_prl_2016}; see numerical evidence in Sec.~S1 of Supplementary~\cite{supp} . Hence, ALTs of $\text{Re}[\epsilon]=0$ occur when the point gap closes, i.e., $\Delta\epsilon=|4\tilde{t}+2\beta(W_c)|$=0. By solving the complex-gap equation, we determine $W_c(\tilde{t})$ for Hamiltonian~\eqref{eq_1} of $d=1$ and plot the phase boundary in Fig.~\ref{fig1}(a), which gives the same $W_c$ obtained by our scaling function Eq.~\eqref{eq_2}. Such consistency strongly support the validity of Eq.~\eqref{eq_2}.
\par

\emph{2D Class AI.}~The one-parameter scaling function Eq.~\eqref{eq_2} is also applicable to higher-dimensional non-reciprocal systems. We now focus on ALTs of $H^{d=2}_{\text{AI}}$. Note that $H^{d=2}_{\text{AI}}$ in the limit of $\tilde{t}\to 0$ belongs to the Gaussian orthogonal ensemble where ALT is prohibited~\cite{fEvers_rmp_2008}. Nonetheless, $H^{d=2}_{\text{AI}}$ of $\tilde{t}\neq 0$ can undergo an ALT. One example is shown in Fig.~\ref{fig2}(e), which depicts $Y_L(W)$ v.s. $W$ for $\text{Re}[\epsilon]=0$, $\tilde{t}=0.1$, and various $L$. One can see all curves cross at $W_c$, as well as a good scaling function $f(x)$ shown in Fig.~\ref{fig2}(f). 
\par

Noticeably, the spectrum of $H^{d=2}_{\text{AI}}$ is gapless, which does not display a point gap like $H^{d=1}_{\text{AI}}$; see Sec.~S1 in Supplementary~\cite{supp}. However, our calculations find that the ALTs mechanisms of the two models are the same, i.e., a state is extended (localized) if its eigenenergy is complex (real), as illustrated by Hatano and Nelson~\cite{nHatano_prl_1996}. Again, we calculate the self-energy of $H^{d=2}_{\text{AI}}$ by using SCBA and determine the critical disorder $W_c$ at which the eigenenergy of the $\text{Re}[\epsilon]=0$ state become real; see Sec.~S3 in Supplementary~\cite{supp}. 
\par

We plot $W_c$ from SCBA in Fig.~\ref{fig1}(d), which is also quantitatively consistent with those from the scaling function Eq.~\eqref{eq_2}. Besides, the obtained critical exponents are independent of $\tilde{t}$, a typical feature of universality as expected; see Fig.~\ref{fig1}(f) and Fig.~S6 in Supplementary~\cite{supp}.
\par

\emph{Class A.}~Symmetries can change the universality of ALTs in non-Hermitian systems~\cite{fEvers_rmp_2008}. For instance, the following non-reciprocal Hamiltonian belongs to a different symmetry class: 
\begin{equation}
\begin{gathered}
H^{d}_{\text{A}}=\sum_{\bm{i}} \left[ c^\dagger_{\bm{i}} (\epsilon_{1,\bm{i}}\sigma_1+\epsilon_{2,\bm{i}}\sigma_2)c_{\bm{i}} \right] \\
+\sum_{\langle \bm{ij}\rangle}\left[ c^\dagger_{\bm{i}}(t+\tilde{t})\sigma_3 c_{\bm{j}}+c^\dagger_{\bm{j}}(t-\tilde{t})\sigma_3 c_{\bm{i}}\right]
\end{gathered}\label{eq_6}
\end{equation}
with $\sigma_{1,2,3}$ being Pauli matrices. The real random numbers $\epsilon_{1,\bm{i}}$ and $\epsilon_{2,\bm{i}}$ distribute uniformly in the range of $[-W/2,W/2]$. Different from model~\eqref{eq_1}, time-reversal symmetry is now broken such that model~\eqref{eq_6} belongs to class A. $H^{d}_{\text{A}}$ for both $d=1$ and $d=2$ can undergo an ALT. Following the same method, we numerically confirm that the scaling function Eq.~\eqref{eq_2} can describe these ALTs and summarize the critical exponents in Table~\ref{table_1}; see Sec.~S2 in Supplementary~\cite{supp}. 
\par

\emph{Non-Hermitian reciprocal systems.}~ALTs also happen at disordered non-Hermitian systems with reciprocity. In our previous works~\cite{cWang_prb_2022,cWang_prb_2023}, we have shown the occurrence of ALTs in 2D classes AIII, DIII+$\mathcal{S}_{-+}$, and DIII+$\mathcal{S}_{-}$, where the hopping terms are symmetric. Here, we further study the ALTs of a random SU(2) model subject to a complex on-site potential including gain and loss, which is reciprocal and belongs to class AII$^\dagger$; see Sec.~S2 in Supplementary~\cite{supp}. The critical exponents, obtained by the scaling function Eq.~\eqref{eq_2}, are given in Table~\ref{table_1}. 
\par

\emph{Discussion.}~The presented scaling function provides a new paradigm for ALTs in non-reciprocal systems, which does not rely on transport properties but on the wavefunctions within the biorthogonal framework~\cite{bi_orthogonal}. This is consistent with the initial ideas of Hanato and Nelson~\cite{nHatano_prb_1998}, that considers a periodic system, rather than an open chain connected to leads. However, the obtained critical exponents $\nu$ are different from those based on the two-parameter scaling theory~\cite{kKawabata_prl_2021}. 
\par

Our numerical calculations summarized in Table~\ref{table_1} sustain that Eq.~\eqref{eq_2} can be applied to non-Hermitian systems belonging to different universality classes. Besides, previous numerical studies show that the criticality of ALTs in disordered Hermitian systems also follows Eq.~\eqref{eq_2}, where $p_2\equiv\left(\sum_{\bm{i}}|\psi^\text{R}_\epsilon(\bm{i})|^4\right)^{-1}$ since $\psi^\text{R}_\epsilon(\bm{i})=\psi^\text{L}_\epsilon(\bm{i})$~\cite{jhPixley_prl_2015,cWang_prb_2019}. From this perspective, we believe that this scaling equation is universal and comprehensive for ALTs. We have not verified that all reciprocal and non-reciprocal non-Hermitian systems of spatial dimensions ($d=1,2,3$) and symmetry classes (38-fold classification) satisfy this scaling function Eq.~\eqref{eq_2}. Nonetheless, this issue deserves further investigations.
\par

\emph{Experimental relevance.}~The non-Hermitian non-reciprocal systems can be implemented in various materials, including electromagnetic~\cite{njPotton_rpp_2004,naEstep_2014,hXu_nature_2019}, magnonic~\cite{tYu_prl_2021}, acoustic~\cite{yWang_prl_2018},  mechanical~\cite{lmNash_pnas_2015},  electronic~\cite{dlSounas_nature_ele_2018}, electric~\cite{thelbig_natphys_2020,dzou_natcommu_2021,ctang_prb_2023}, and quantum systems~\cite{vPeano_prx_2016}, while the disorder effect is pervasive. However, manually regulating the strength of randomness to verify the scaling equation remains challenging. Here, we present a proposal in electric circuits that can directly measure the participation ratio and test the correctness of Eq.~\eqref{eq_2}; see Sec.~S4 Supplementary~\cite{supp}.
\par

\emph{Summary.}~We propose a one-parameter scaling function for ALTs in non-reciprocal non-Hermitian systems and prove its validity in four universality classes. Besides, we derive a complex-gap equation to determine the critical disorder $W_c$ based on SCBA since ALTs in these systems share similarities from a topological perspective, i.e., ALTs occur when eigenenergies become real. The obtained critical disorders match those from the scaling function Eq.~\eqref{eq_2}. We also apply Eq.~\eqref{eq_2} to some disordered non-Hermitian reciprocal systems; see Table~\ref{table_1}. Therefore, the central question addressed in this article is whether there exists a 1PST for non-Hermitian (especially, non-reciprocal) localization, and our answer is positive such that one can understand a large class of Anderson localization problems within a unified theoretical framework.
\par

\begin{acknowledgments}

This work is supported by the National Natural Science Foundation of China (Grants No.~11774296, No.~11704061, and No.~11974296) and Hong Kong RGC (Grants No.~16300522 and No.~16302321).

\end{acknowledgments}


\begin{thebibliography}{99}
\bibitem{bi_orthogonal}
The right and left eigenstates of a non-Hermitian Hamiltonian are defined by: $H|\psi^{\text{R}}_{\epsilon}\rangle=\epsilon |\psi^{\text{R}}_{\epsilon}\rangle$ and $H^\dagger |\psi^{\text{L}}_{\epsilon}\rangle=\epsilon^\ast |\psi^{\text{L}}_{\epsilon}\rangle$. The right and left eigenstates follow the biorthogonal relation: $\langle \psi^{\text{L}}_{\epsilon'}|\psi^{\text{R}}_{\epsilon}\rangle=\delta_{\epsilon,\epsilon'} $.
\bibitem{numerical}
We numerically compute $p_2(\epsilon,W)$ by using the KWTANT package~\cite{kwant} and the SciPy library~\cite{scipy} on Python.
\end{thebibliography}
\end{document}